\documentclass[conference]{IEEEtran}

\usepackage{amsmath,amssymb,amsfonts}
\usepackage{graphicx}
\usepackage{cite}
\usepackage{booktabs}
\usepackage{bm}
\usepackage{array}

\begin{document}
	\title{Network-Verified NTN Positioning for 6G: A Standards-Oriented Survey of Hybrid TN--NTN Localization}
	\author{\IEEEauthorblockN{Donglin Wang\IEEEauthorrefmark{1}, Zexing Fang\IEEEauthorrefmark{1}, Qiuheng Zhou\IEEEauthorrefmark{1}, and Hans D. Schotten\IEEEauthorrefmark{1}\IEEEauthorrefmark{2}}
		\IEEEauthorblockA{\IEEEauthorrefmark{1}\textit{Rhineland-Palatinate Technical University of Kaiserslautern-Landau, Germany}\\
		\{donglin.wang, zfang\}@rptu.de}
		\IEEEauthorblockA{\IEEEauthorrefmark{2}\textit{German Research Center for Artificial Intelligence (DFKI GmbH), Kaiserslautern, Germany}\\
		\{qiuheng.zhou, schotten\}@dfki.de}
	}
	\maketitle

\begin{abstract}
Wireless positioning is evolving from legacy cellular and standalone satellite navigation toward hybrid, high-accuracy, three-dimensional (3D), and network-verifiable frameworks for 5G-Advanced, 6G, and non-terrestrial network (NTN) systems. This shift is driven by intelligent transportation, unmanned aerial vehicle (UAV) operation, low-altitude economy services, public warning, and regulated NTN use cases that require not only accurate location estimates, but also vertical awareness and verification confidence. This survey reviews the evolution from Long Term Evolution (LTE) and Global Navigation Satellite System (GNSS) positioning to New Radio (NR), Release-18/19 positioning enhancements, and NTN-enabled 3D positioning. It synthesizes observable families, 3rd Generation Partnership Project (3GPP) standard evolution, method-level tradeoffs, and open challenges for hybrid terrestrial network (TN)--NTN designs, with emphasis on standardization impact, vertical observability, reliability-aware measurement selection, and network-side verification.
\end{abstract}


\section{Introduction}
Positioning has become a core capability of modern wireless systems because many services now depend on accurate, timely, and trustworthy location information. Navigation and emergency response are increasingly joined by intelligent transportation, unmanned aerial vehicle (UAV), and low-altitude mobility, asset tracking, public warning, industrial automation, and context-aware communication. These applications push positioning beyond horizontal accuracy toward vertically aware, hybrid, and verification-oriented operation across heterogeneous environments.

In this paper, positioning and localization are used to denote the estimation of a UE's location from radio, satellite, or sensor measurements; 3D positioning denotes estimation of both horizontal coordinates and altitude; and network-verified location denotes a network-side assessment of whether a UE-reported location is consistent with independently available measurements, constraints, and confidence bounds. Trustworthiness therefore refers not only to small error, but also to confidence, integrity risk, and the ability to reject unreliable or manipulated reports.

The 3rd Generation Partnership Project (3GPP) standardization trajectory reflects this broader role. In Technical Specification (TS) 38.305 for the Next Generation Radio Access Network (NG-RAN), user equipment (UE) positioning supports or assists the calculation of a UE's geographical position, with estimation performed either by the UE or by the Location Management Function (LMF) \cite{TS38305}. TS~38.305 also explicitly supports hybrid positioning using multiple methods \cite{TS38305}, which is essential because no single method can simultaneously guarantee coverage, accuracy, blockage robustness, low signaling cost, and trustworthiness in all scenarios.

This survey interprets the evolution in three steps. First, Long Term Evolution (LTE) established the cellular positioning framework through network-assisted Global Navigation Satellite System (GNSS), downlink positioning, enhanced Cell ID, uplink positioning, and short-range technologies such as wireless local area network (WLAN) and Bluetooth \cite{TS36305}. Second, GNSS and assisted-GNSS (A-GNSS) remain a dominant outdoor baseline because of their global reach and native three-dimensional (3D) capability \cite{ETSI_GNSS}. Third, New Radio (NR), 5G-Advanced, and non-terrestrial network (NTN) operation extend positioning toward multi-method fusion, carrier-phase refinement, sidelink positioning, artificial intelligence/machine learning (AI/ML)-assisted reliability handling, vertical observability, and network-verified location \cite{TS38305,TR38843,TS38355}. In particular, Technical Report (TR) 38.882 highlights the need to verify UE-reported location in NTN for regulated services such as emergency calls, lawful intercept, public warning, and charging or billing \cite{TR38882}.

The central perspective is therefore not only a transition from LTE to NR or from terrestrial to non-terrestrial access. It is a transition from isolated mechanisms to integrated frameworks in which observables, infrastructure domains, geometry, signaling, and trust models must be considered jointly. Compared with prior cellular or 6G localization surveys, this paper emphasizes the standards-to-observable path toward hybrid 3D and network-verified TN--NTN positioning, which directly matches the need for standards-aware research on communications and networking evolution.

The survey makes four contributions:
\begin{itemize}
    \item It provides a standards-to-observable mapping from LTE and GNSS baselines to NR, 5G-Advanced Release-18/19, sidelink, and NTN-oriented positioning.
    \item It organizes timing, phase, angle, signal-strength, GNSS, Doppler, and hybrid measurements by synchronization needs, dominant impairments, and 3D observability.
    \item It distinguishes location estimation from network-side verification and connects this distinction to NTN regulatory, privacy, and mission-critical use cases.
    \item It identifies research gaps for hybrid TN--NTN positioning, including vertical observability, reliability-aware node selection, integrity, and benchmarkable verification metrics.
\end{itemize}

The remainder of this paper is organized as follows. Section~II reviews the state of the art and clarifies the survey scope. Section~III introduces the method taxonomy, including GNSS and A-GNSS. Section~IV reviews standard evolution and the method-to-standard mapping. Sections~V--VII review LTE, NR, and NTN positioning methods. Section~VIII compares the main paradigms, Section~IX discusses open challenges and research opportunities, Section~X gives the future outlook and design guidelines, and Section~XI concludes the paper.

\section{State of the Art and Survey Scope}
The IEEE literature places the standards evolution in context. Prior surveys cover cellular positioning from 1G to 5G \cite{DelPeral2018Survey}, tutorial-level 5G positioning architectures and algorithms \cite{Italiano2025Tutorial}, broader 6G localization trends \cite{Trevlakis2023Localization}, ground-air-space localization \cite{Sallouha2025GAS}, vehicular wireless positioning \cite{Saleh2026Vehicular}, and reconfigurable intelligent surface (RIS)-aided radio localization \cite{Umer2025RIS}. Foundational work on cooperative localization and network navigation shows how cooperation among nodes can improve localization reliability \cite{Wymeersch2009Cooperative,Win2011Cooperation}, while 5G millimeter-wave (mmWave) studies show that wide bandwidths and large antenna arrays can support high-resolution vehicular positioning and joint position-orientation estimation \cite{Wymeersch2017MmWave,Shahmansoori2018MmWave}. More focused work addresses NR positioning studies and Release-18 enhancements \cite{TR38855,TR38857,TR38859,Cha2025Rel18}, carrier-phase indoor accuracy improvements \cite{Nikonowicz2024Indoor}, and NTN or satellite-enabled positioning visions \cite{TR38811,Dureppagari2023NTN,DelPeral2024Satellite}.

These works motivate, but also distinguish, the scope of this survey: rather than another general localization overview, it follows the standards-aware path from LTE/GNSS baselines to NR/NTN-enabled 3D positioning, emphasizing standardized observables, TN--NTN hybrid fusion, vertical observability, selective measurement usage, and network-side verification of UE-reported location. In short, the contribution is a compact standards-to-observable synthesis for CSCN-style discussion, not an exhaustive algorithm catalogue.

The state of the art is moving from method-specific localization toward hybrid and verification-aware frameworks. Earlier surveys emphasized cellular positioning evolution, while recent work treats multi-round trip time (Multi-RTT), time difference of arrival (TDOA), angle of arrival (AoA), GNSS, carrier phase, millimeter-wave (mmWave), and sensor inputs as components of a network-coordinated architecture \cite{DelPeral2018Survey,Italiano2025Tutorial,Trevlakis2023Localization,Wymeersch2017MmWave,Shahmansoori2018MmWave,Fan2022CarrierPhase,Nikonowicz2024Indoor}.

This shift is consistent with TS~38.305, which formalizes NG-RAN positioning methods, LMF coordination, assistance delivery, and hybrid operation \cite{TS38305}. It is also supported by NR positioning studies in Releases 16--18, Release-19 AI/ML work for NR air-interface positioning, sidelink positioning protocol support, and positioning signaling specifications such as LTE Positioning Protocol (LPP) and NR Positioning Protocol A (NRPPa) \cite{TR38855,TR38857,TR38859,TR38843,TS38355,TS37355,TS38455}. TR~38.882 adds the NTN verification dimension: the network must assess location independently of UE-reported values while remaining compatible with existing NG-RAN procedures and radio access technology (RAT)-dependent methods \cite{TR38882}.

The literature can be grouped into cellular surveys, high-accuracy phase-based positioning, NR standardization, mmWave/cooperative localization, NTN and satellite-assisted positioning, and reliability-aware node selection. Across these streams, the remaining gap is the joint treatment of vertical observability, TN--NTN fusion, and network-side verification.

Two gaps remain especially visible. First, NTN positioning is still dominated by vision papers and system-level studies rather than mature, benchmarked end-to-end solutions. Second, robust altitude estimation under blockage, heterogeneous synchronization, and TN--NTN geometry diversity remains underexplored, even though it is central to UAV, low-altitude economy, and network-verified NTN scenarios.

\section{Background and Taxonomy of Positioning Methods}
\begin{figure*}[!t]
\centering
\includegraphics[width=0.55\textwidth]{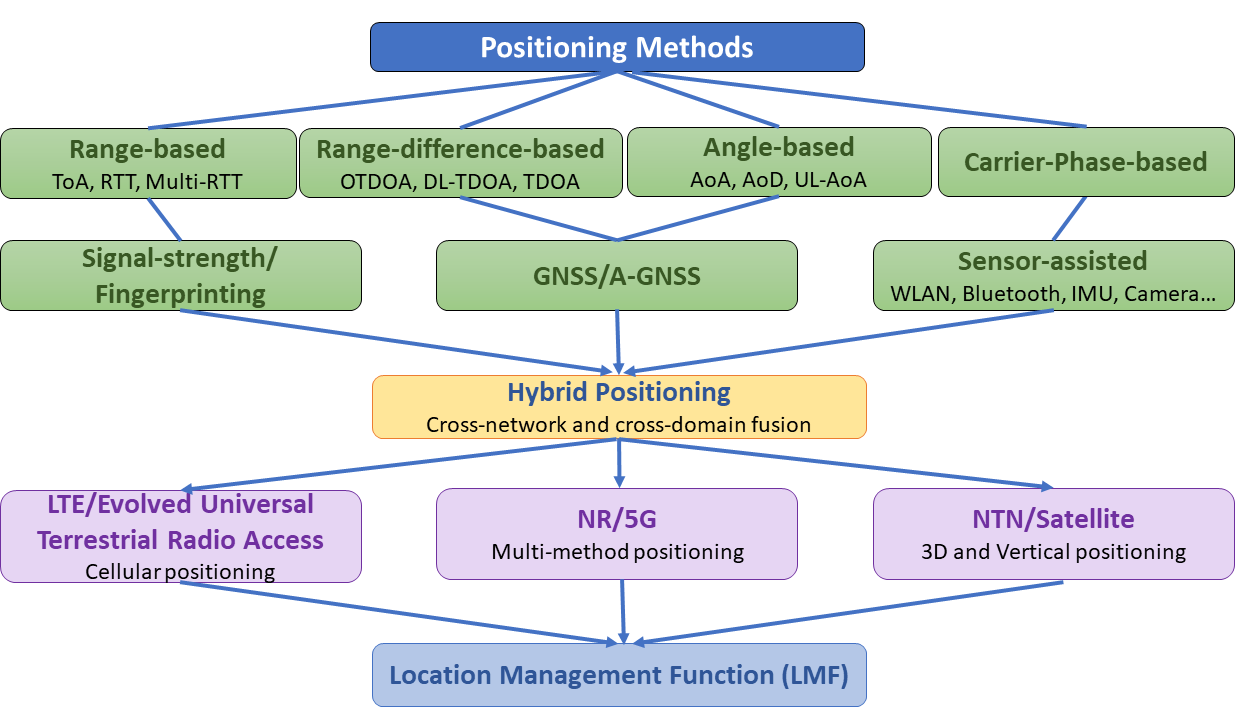}
\caption{Taxonomy of positioning methods from timing, phase, angle, and other observables to hybrid LTE, NR, and NTN-enabled positioning.}
\label{fig:taxonomy}
\end{figure*}

Positioning methods can be classified by the information they exploit: propagation time, time difference, carrier phase, angle, received power, stored radio fingerprints, satellite navigation data, Doppler/frequency shift, or fused multi-source measurements \cite{DelPeral2018Survey,Italiano2025Tutorial,Trevlakis2023Localization}. This measurement-domain taxonomy is complemented by architecture choices, namely UE-based, UE-assisted, network-based, and hybrid estimation. The distinction matters because each class has different synchronization assumptions, infrastructure requirements, dominant error sources, 3D observability, and verification potential. Fig.~\ref{fig:taxonomy} summarizes the taxonomy used in this survey, from basic observables to cross-domain hybrid fusion.

\subsection{Range-Based Methods}
Range-based methods estimate distance from propagation delay. One-way time of arrival (ToA) requires tight transmitter-receiver clock knowledge, whereas two-way round-trip time (RTT) and Multi-RTT reduce UE clock dependence at the cost of extra signaling and processing-delay calibration \cite{TS38305,TR38855,Italiano2025Tutorial,Sinha2022TOA}. A measured delay is converted into distance after accounting for clock, hardware, and protocol offsets, so accuracy improves with bandwidth, line-of-sight (LoS) availability, and stable synchronization. The main limitations are clock bias, non-line-of-sight (NLoS) propagation, timing-advance quantization, and poor anchor geometry, which can turn precise timing into biased position estimates.

\subsection{Range-Difference-Based Methods}
Range-difference-based methods use differences rather than absolute distances. A representative example is TDOA, where the location is inferred from hyperbolic constraints created by relative arrival times at multiple transmission or reception points \cite{Fang2025Lightweight,Motie2024SelfUAV}. This does not correct the UE clock itself; instead, subtracting measurements with respect to a reference node cancels the common UE clock term in the TDOA equations. The cost is that at least one additional independent anchor/reference is needed, and the synchronization burden shifts to the network. TDOA also amplifies geometry-dependent error propagation when anchors are nearly collinear or poorly distributed. Representative standardized variants include observed TDOA (OTDOA), downlink (DL) TDOA (DL-TDOA), and uplink (UL) TDOA (UL-TDOA) \cite{TS36305,TS38305}; their accuracy depends on inter-site synchronization, reference-signal detectability, and geometric dilution of precision.

\subsection{Angle-Based Methods}
Angle-based methods estimate the AoA or angle of departure (AoD) of signals by exploiting antenna arrays. In NR, angle information becomes increasingly relevant because massive multiple-input multiple-output (MIMO), beam management, and higher carrier frequencies make directional measurements more practical \cite{Wymeersch2017MmWave,Shahmansoori2018MmWave}. Angle measurements improve horizontal and vertical observability, especially when zenith angles are available and fused with timing measurements; representative standardized variants include DL-AoD and UL-AoA \cite{TS38305}. The main challenges are array calibration, angular ambiguity, beam-training overhead, multipath reflections, LoS dependence, and reduced resolution when aperture or signal-to-noise ratio is limited.

\subsection{Signal-Strength and Fingerprinting Methods}
These methods use received signal strength, reference-signal received power, or radio fingerprints constructed from previously observed signal maps \cite{DelPeral2018Survey,Italiano2025Tutorial}. Signal-strength methods are simple and can provide coarse localization when timing or angle measurements are unavailable, but path-loss variability limits accuracy in complex environments. Fingerprinting improves robustness by matching online measurements to a database of cellular, WLAN, Bluetooth, or sensor patterns, including auxiliary WLAN and Bluetooth positioning support in LTE/NR frameworks \cite{TS36305,TS38305}. Their main drawbacks are database maintenance, calibration effort, environmental change, and device-dependent measurement behavior.

\subsection{GNSS-Based Methods}
GNSS-based methods rely on satellite signals to compute position, velocity, and time-related navigation quantities \cite{ETSI_GNSS}. They are a widely deployed outdoor baseline, with wide-area coverage, mature chipset support, and native 3D capability through pseudorange and, in high-accuracy modes such as RTK/PPP, carrier-phase measurements. Assisted GNSS (A-GNSS) improves acquisition time and sensitivity by delivering assistance data, timing, and support information through the cellular network, while the UE contributes GNSS measurements or estimates \cite{TS36305,TS38305}. GNSS is also valuable as an independent reference for network-verified positioning, but it can degrade indoors, in dense urban canyons, under blockage, or under interference, spoofing, and jamming \cite{TR38882,DelPeral2024Satellite}.

\subsection{Phase-Based and Carrier-Phase Methods}
Carrier-phase positioning (CPP) uses received carrier phase rather than only code or envelope timing. Its sub-wavelength precision makes it a key 5G-Advanced observable for centimeter-level indoor or dense-network positioning \cite{Fan2022CarrierPhase,Nikonowicz2024Indoor,Cha2025Rel18}. The main difficulty is carrier-cycle ambiguity: integer wavelengths must be resolved or tracked, and phase is sensitive to oscillator instability, phase noise, NLoS propagation, and cycle slips. CPP is therefore most useful when fused with timing, angle, inertial, or GNSS measurements that constrain ambiguity search and detect unreliable links.

\subsection{Doppler- and Frequency-Based Methods}
Doppler and frequency-offset measurements are especially relevant for LEO NTN links, where satellite motion produces large and rapidly varying frequency shifts \cite{TR38811,Dureppagari2023NTN}. When satellite ephemeris and oscillator behavior are known, Doppler can constrain radial velocity and help disambiguate timing measurements; when they are uncertain, it becomes an additional error source. Doppler-aided positioning therefore requires joint modeling of UE motion, satellite dynamics, clock drift, and carrier-frequency offset, and is most useful as a complementary NTN observable rather than a standalone method.

\subsection{Hybrid Methods}
Hybrid methods fuse information from GNSS, cellular timing, angle measurements, inertial measurement units (IMUs), barometers, WLAN, Bluetooth, or other auxiliary sources to improve robustness, accuracy, and coverage \cite{TS38305,Italiano2025Tutorial,Trevlakis2023Localization}. Fusion may combine raw observables or weight independent position estimates according to reliability. Hybridization is especially important for 3D and TN--NTN positioning because no single source is reliable across all propagation, mobility, and visibility conditions \cite{TR38882,Dureppagari2023NTN}. Practical fusion can use weighted least squares for snapshot estimation, Kalman filtering for tracking, factor graphs for asynchronous multi-sensor constraints, or artificial intelligence (AI)-assisted reliability estimation for non-line-of-sight and link-quality classification \cite{Wymeersch2009Cooperative,Win2011Cooperation,Italiano2025Tutorial}. The key design problem is therefore how to select, weight, and verify measurements under changing geometry and signal quality \cite{Fang2025Lightweight,TS37355,TS38455}.

\subsection{Compact Estimation View}
Most positioning methods can be expressed through a common model that shows why observable choice, node geometry, and measurement quality affect localization accuracy. Let $z_i$ denote the $i$-th observed timing, angle, Doppler, GNSS, or carrier-phase measurement, and let $h_i(\bm{x})$ be its prediction from the UE state $\bm{x}$, which includes position and nuisance states such as clock bias or carrier-phase ambiguity. For a selected measurement set $\mathcal{S}$, define the stacked residual noise $\bm{n}_{\mathcal{S}}=\bm{z}_{\mathcal{S}}-\bm{h}_{\mathcal{S}}(\bm{x})$, the Jacobian $\bm{H}_{\mathcal{S}}=\partial\bm{h}_{\mathcal{S}}/\partial\bm{x}$, and the noise covariance $\bm{R}_{\mathcal{S}}=\mathbb{E}[\bm{n}_{\mathcal{S}}\bm{n}_{\mathcal{S}}^{T}]$. Then
\begin{equation}
\hat{\bm{x}}=\arg\min_{\bm{x}}\sum_{i\in \mathcal{S}} w_i\|z_i-h_i(\bm{x})\|^2,\qquad
\bm{J}_{\mathcal{S}}=\bm{H}_{\mathcal{S}}^{T}\bm{R}_{\mathcal{S}}^{-1}\bm{H}_{\mathcal{S}},
\end{equation}
where $w_i$ is a reliability weight. The information matrix $\bm{J}_{\mathcal{S}}$ determines the local Cram\'er--Rao lower bound (CRLB), with position-error covariance bounded by $\bm{J}_{\mathcal{S}}^{-1}$ under the linearized unbiased model. Thus, poor geometry weakens $\bm{H}_{\mathcal{S}}$, while noisy or biased links enlarge $\bm{R}_{\mathcal{S}}$; both directly degrade localization accuracy and verification confidence.

Representative observation terms include $z_i^{\rm ToA}\!\approx\!\|\bm{p}-\bm{a}_i\|/c+b$, $z_{ij}^{\rm TDOA}\!\approx\!(\|\bm{p}-\bm{a}_i\|-\|\bm{p}-\bm{a}_j\|)/c$, $z_i^{\rm CPP}\!\approx\!\|\bm{p}-\bm{a}_i\|+N_i\lambda$, and $z_i^{\rm AoA}\!\approx\!\angle(\bm{p}-\bm{a}_i)$, where $\bm{p}$ is UE position, $\bm{a}_i$ is an anchor, $b$ is clock bias, $N_i$ is an integer ambiguity, and $\lambda$ is wavelength. This form makes the later discussion of CRLB/GDOP, reliability-aware node selection, and verification residuals a consequence of the measurement model rather than a separate assumption.

\subsection{3D Observability and Vertical Positioning}
3D positioning is not obtained simply by adding an altitude coordinate to a 2D solver. Vertical accuracy is often worse than horizontal accuracy because terrestrial anchors are nearly coplanar, producing weak elevation diversity and poor vertical dilution of precision. Under limited elevation diversity, solver choice also matters: linear least-squares (LS) or algebraic TDOA reformulations are simple but can suffer from linearization or reformulation errors, whereas nonlinear solvers such as gradient descent, Gauss--Newton, or Levenberg--Marquardt optimize the original range, TDOA, angle, or phase model and may outperform linear LS when initialization and outlier handling are adequate \cite{Italiano2025Tutorial}. Zenith-angle measurements, barometers, inertial sensors, GNSS altitude, and carrier-phase constraints can improve altitude estimates, but each has failure modes such as calibration drift, blockage, ambiguity, or NLoS bias. TN--NTN fusion can add non-coplanar geometry, but satellite-based aiding is not a free extra anchor because ephemeris, Doppler, beam, and timing assistance may be protocol-heavy. Robust 3D positioning therefore needs selective geometry-aware anchor activation, suitable nonlinear or hybrid solvers, altitude-specific metrics, and integrity checks.

\section{Standard Evolution from LTE to NR and NTN}
\begin{figure}[htp]
\centering
\includegraphics[width=0.90\linewidth]{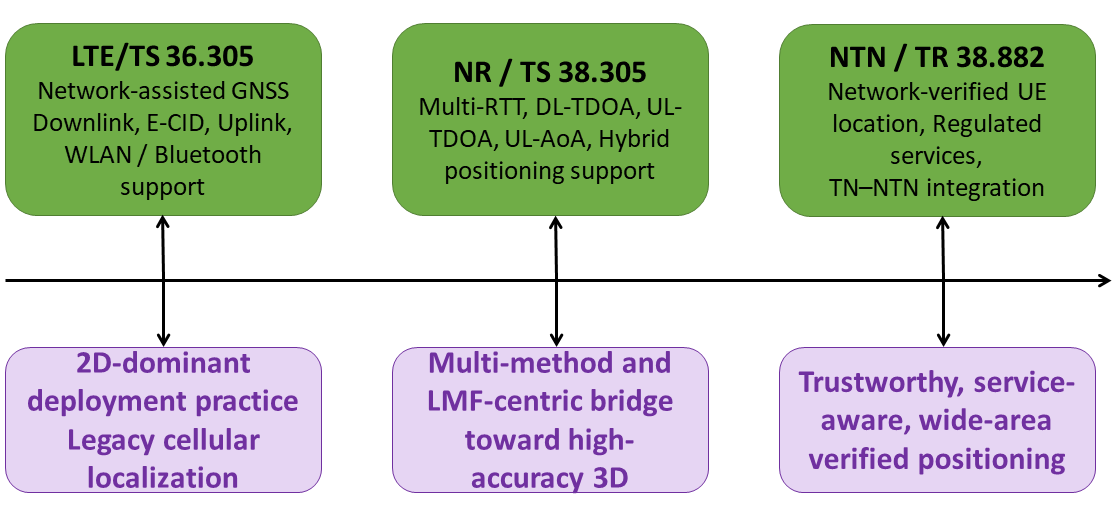}
\caption{Standard evolution from LTE positioning to NR multi-method positioning and NTN-oriented network-verified UE location.}
\label{fig:standards_evolution}
\end{figure}

Fig.~\ref{fig:standards_evolution} summarizes the chronological shift from LTE positioning to NR multi-method positioning and NTN-oriented verification. The key standards distinction is functional role: TS~36.305 and TS~38.305 define stage-2 positioning architecture for E-UTRAN and NG-RAN, TS~37.355 and TS~38.455 define LPP/NRPPa signaling, and TR~38.855/857/859 document the NR study path from Release 16 to Release 18 \cite{TS36305,TS38305,TS37355,TS38455,TR38855,TR38857,TR38859}. Release 19 continues this path through updated NG-RAN positioning, AI/ML positioning accuracy enhancements, and the NR Sidelink Positioning Protocol (SLPP), which broadens the standards discussion from infrastructure-only localization to UE-to-UE and model-assisted operation \cite{TR38843,TS38355}. NTN-related work does not replace terrestrial methods; TR~38.882 reuses existing RAT-dependent procedures as baselines for network-verified UE location in regulated services \cite{TR38882}.

\subsection{Method-to-Standard Mapping}
The method-to-standard mapping is best read as accumulation rather than replacement. LTE provides the cellular baseline; NR adds denser multi-node timing, angle, sensor-assisted, and hybrid operation; 5G-Advanced studies elevate carrier phase as a high-accuracy cellular observable; Release 19 adds AI/ML-assisted and sidelink dimensions; and NTN shifts the emphasis from estimation alone to verification under wide-area coverage and regulatory constraints \cite{TS36305,TS38305,Fan2022CarrierPhase,Nikonowicz2024Indoor,Cha2025Rel18,TR38843,TS38355,TR38882}. Thus, the main standardization trend is coordinated use of existing observables through LMF, LPP, NRPPa, and SLPP rather than a single new positioning algorithm.

\section{LTE Positioning Methods}
LTE TS~36.305 provides a multi-method framework rather than a single localization algorithm. The estimate may be computed at the UE or by the evolved Serving Mobile Location Centre (E-SMLC), using network-assisted GNSS, downlink OTDOA, enhanced Cell ID (E-CID), uplink measurements, WLAN, Bluetooth, or combinations of these sources \cite{TS36305}.

\subsection{A-GNSS and OTDOA}
A-GNSS is the strongest LTE outdoor baseline: the network supplies assistance data, coarse timing, and satellite information, while the UE measures or computes the GNSS solution. Its limitations are the familiar GNSS ones: indoor attenuation, urban-canyon multipath, interference, spoofing, and weak altitude reliability under poor satellite geometry. OTDOA complements GNSS by using observed time differences between positioning reference signals from multiple eNBs. It depends on inter-cell synchronization, PRS hearability, bandwidth, and favorable geometry; wider PRS bandwidth, repeated occasions, and muting can improve timing resolution and interference control, but NLoS delay bias and weak cell geometry often dominate indoors or in sparse macro layouts.

\subsection{E-CID, Uplink, and Local-Area Aiding}
E-CID refines cell identity using timing advance, received signal strength/quality, and related radio measurements. It has low overhead and broad availability, but timing-advance granularity and cell size limit accuracy. Uplink positioning shifts measurements to the network side, which can be useful for verification, but it requires suitable receiving points, calibration, and synchronization. WLAN and Bluetooth support can improve indoor coverage through proximity or fingerprinting, at the cost of database maintenance, device dependence, and environmental drift. Thus, LTE regimes are asymmetric: A-GNSS is strongest in open sky, OTDOA requires multiple synchronized and hearable cells, E-CID is robust but coarse, uplink methods support network-side control, and WLAN/Bluetooth mainly help indoors. LTE can support 3D estimates in principle, but practical deployments are often horizontal-dominant because vertical accuracy needs elevation diversity, barometric/sensor aiding, or hybrid GNSS-cellular fusion.

From a measurement perspective, E-CID mainly narrows the UE to a serving-cell sector or timing-advance ring, while uplink methods can add network-observed time or angle constraints when receiving points are calibrated. This makes LTE useful as a fallback and assistance layer, but not a complete answer for vertical or integrity-sensitive services.

\section{NR as the Bridge to High-Accuracy and 3D-Aware Positioning}
NR materially improves the LTE baseline through larger bandwidth, denser deployments, beamforming, flexible PRS/SRS resources, and LMF-centered coordination. Wider bandwidth improves delay resolution, dense small cells improve geometry, beamforming exposes angular observables, and configurable reference resources allow finer control over bandwidth, periodicity, repetition, and muting. TS~38.305 supports UE-based, UE-assisted, LMF-based, and NG-RAN-node-assisted operation, enabling timing, angle, carrier-phase, sensor-assisted, and hybrid methods to be requested and fused according to service requirements \cite{TS38305}.

\subsection{Timing-Based Methods}
Multi-RTT, DL-TDOA, and UL-TDOA all exploit timing, but their assumptions differ. Multi-RTT estimates ranges through two-way exchanges with multiple transmission/reception points, reducing strict UE-network clock alignment requirements but increasing signaling and delay-calibration needs. DL-TDOA uses downlink reference-signal timing differences observed by the UE, so inter-gNB synchronization and PRS detectability are critical. UL-TDOA moves timing observations to the network, enabling network-side consistency checks but requiring synchronized receivers and sufficient uplink geometry \cite{TS38305,TR38855,TR38857}. A practical NR deployment may select among them according to latency, UE capability, privacy policy, and whether the service needs estimation or verification.

\subsection{Angle-Based Methods}
NR angle methods use antenna arrays and beamformed measurements. DL-AoD constrains the UE from the departure direction of downlink beams, while UL-AoA estimates the arrival direction of UE transmissions at the network. Azimuth and zenith angles are especially valuable for 3D positioning because they add vertical observability that pure terrestrial timing can lack. Their practical limits are array calibration, aperture, beam training overhead, LoS/NLoS discrimination, and multipath angular bias \cite{Wymeersch2017MmWave,Shahmansoori2018MmWave}.

\subsection{Carrier-Phase Enhancements}
Carrier-phase measurements add sub-wavelength resolution and are central to 5G-Advanced high-accuracy positioning, but require ambiguity resolution, phase stability, and cycle-slip detection \cite{Fan2022CarrierPhase,Nikonowicz2024Indoor,Cha2025Rel18}. They should be treated as a refinement layer: timing and angle measurements constrain ambiguity search, sensors detect motion consistency and cycle slips, and GNSS or network geometry provides external checks.

\subsection{Sensor-Assisted and Hybrid Methods}
Sensor-assisted and hybrid positioning combine timing, angle, GNSS, inertial sensors, barometers, WLAN/Bluetooth, and terrestrial beacons. In practice, sensors smooth motion between radio updates, barometers add altitude cues, and GNSS or WLAN/Bluetooth can provide independent checks when cellular geometry is weak. The main challenge is reliability weighting: stale fingerprints, drifting inertial sensors, or NLoS radio links should not be fused as equally trustworthy measurements.

\subsection{3D Positioning and Release-18/19 Implications}
Release 18-oriented enhancements strengthen NR through better multi-method operation, improved accuracy targets, carrier-phase support, and tighter LMF coordination \cite{TR38859,Cha2025Rel18}. Release 19 extends the standards path toward AI/ML-assisted positioning and sidelink positioning, where learned line-of-sight or measurement-quality indicators and UE-to-UE constraints can complement infrastructure measurements \cite{TR38843,TS38355}. The 3D gain comes from combining timing ranges, DL-AoD/UL-AoA zenith information, carrier-phase refinement, sidelink or sensor constraints, and altitude cues rather than from any single method. This bridges LTE-style method support toward high-accuracy and verification-capable positioning.

\section{NTN Positioning and Network-Verified UE Location}
NTN introduces satellites, high-altitude platform stations (HAPS), and other aerial platforms into the cellular ecosystem. TR~38.811 studies NR support for NTN, while TR~38.882 studies network-verified UE location for NTN in NR \cite{TR38811,TR38882}.

\subsection{Assistance, Estimation, and Verification}
Network-assisted UE positioning provides assistance data and lets the UE compute or support the estimate. Network-estimated positioning moves more computation and measurement collection to the network. Network-verified UE-reported positioning is different: the network checks whether a UE-reported location is consistent with independent measurements, confidence bounds, service area constraints, and regulatory requirements. This distinction is crucial for emergency calls, lawful intercept, public warning, charging, and cross-border access control, where accuracy alone is insufficient \cite{TR38882}.

\subsection{NTN-Specific Measurement Challenges}
Applying terrestrial baselines to NTN links is non-trivial. LEO mobility creates large Doppler shifts and fast geometry changes; long propagation paths increase RTT, timing-advance ambiguity, and delay calibration sensitivity; ephemeris and satellite-clock errors bias range observables; moving beams and large beam footprints create cell-identity ambiguity; sparse/nonuniform geometry weakens dilution of precision; and mixed TN--NTN deployments introduce synchronization heterogeneity \cite{TR38811,Dureppagari2023NTN,DelPeral2024Satellite}. For timing methods these effects enter as clock, propagation, and ephemeris terms; for angle methods, long ranges and moving beams reduce angular discrimination unless terrestrial anchors are fused.

\subsection{Candidate Verification Solutions}
GNSS-assisted checks provide an independent outdoor reference when available; TN--NTN fusion adds complementary geometry; Doppler-aided correction can reduce timing ambiguity; and consistency checks across GNSS, timing, angle, beam identity, and UE reports can expose unreliable claims. Since no single NTN platform simultaneously provides independent, geometrically diverse, and synchronization-compatible constraints, NTN verification must be hybrid by design. Integrity monitoring and residual-based reliability filtering should attach confidence and fault indicators, while privacy-aware protocols may return a bounded area, service-area pass/fail decision, or integrity flag rather than exact coordinates. NTN positioning is therefore not only a coverage problem, but also a trust, privacy, and regulatory-confidence problem.
Platform type changes which verification function is stressed: LEO links make reliability screening hardest because ephemeris freshness, Doppler, and beam motion change rapidly; HAPS links offer more stable geometry but still require beam-footprint ambiguity checks; and UAV relays need stronger integrity flags because relay motion and calibration can invalidate historical residuals.

For regulatory use, the network may only need to confirm that the UE is inside an allowed country, beam footprint, or emergency-service area. This motivates confidence-bounded verification outputs and auditability, while avoiding unnecessary exposure of precise user coordinates.

\section{Comparative Discussion}
Table~\ref{tab:method_compare} gives a normalized qualitative comparison; exact accuracy and latency remain deployment-dependent.

\begin{table*}[!t]
\caption{Structured comparison of representative positioning method families.}
\label{tab:method_compare}
\centering
\scriptsize
\setlength{\tabcolsep}{3pt}
\renewcommand{\arraystretch}{0.9}
\resizebox{\textwidth}{!}{%
\begin{tabular}{lllll}
\toprule
\textbf{Family} & \textbf{Observable} & \textbf{Key requirements} & \textbf{Dominant limitations} & \textbf{3D / NTN / verification role} \\
\midrule
GNSS/A-GNSS & Pseudorange, Doppler, carrier phase & Satellite visibility; assistance; interference control & Blockage, urban canyon, spoofing/jamming & Outdoor 3D baseline and verification reference; weak indoors \cite{ETSI_GNSS,TS36305,TR38882} \\
ToA/RTT/Multi-RTT & Absolute / two-way time & Bandwidth; delay calibration; multiple anchors & Clock bias, NLoS, processing delay, overhead & NR ranging; 3D needs elevation diversity; NTN needs delay correction \cite{TS38305,TR38855} \\
TDOA variants & Relative arrival time & Network sync; detectable reference signals; geometry & Timing errors, GDOP, weak anchors & Common UE-clock cancellation; multi-anchor consistency checks; NTN sync adaptation \cite{TS36305,TS38305,Fang2025Lightweight} \\
AoA/AoD & Bearing / zenith angle & Arrays; calibration; beam management; LoS/multipath resolution & Aperture, beam overhead, ambiguity, multipath & Altitude via zenith angle; NTN angular diversity needs TN fusion \cite{Wymeersch2017MmWave,Shahmansoori2018MmWave} \\
CPP & Carrier phase & Phase stability; ambiguity and cycle-slip handling & Integer ambiguity, phase noise, NLoS phase bias & Centimeter refinement and residual checks when reliable \cite{Fan2022CarrierPhase,Nikonowicz2024Indoor} \\
E-CID/RSS/fingerprinting & Cell ID, power, maps & Dense infrastructure or maintained database & Coarse accuracy, drift, device dependence & Low-overhead fallback; weak for precise 3D/regulatory NTN \cite{TS36305,TS38305} \\
Hybrid TN--NTN & Multi-observable fusion & LMF coordination; reliability weights; integrity monitoring & Complexity, overhead, inconsistent timing/trust & Best route to robust 3D and network-verified positioning \cite{TR38882,Dureppagari2023NTN,DelPeral2024Satellite} \\
\bottomrule
\end{tabular}
}
\end{table*}

The comparison shows four transitions: LTE/GNSS baselines give way to NR multi-node and multi-observable positioning; horizontal-dominant deployments evolve toward 3D estimation; isolated methods are replaced by hybrid fusion; and NTN changes the target from accuracy alone to verified, integrity-aware location. The tradeoff is complexity: stronger 3D and trust support requires tighter synchronization, ambiguity handling, reliability weighting, privacy control, and LMF coordination \cite{TS36305,TS38305,TR38882,Fang2025Lightweight}.

\section{Open Challenges and Research Opportunities}
Five challenges remain central. First, vertical observability is weak under coplanar anchors or GNSS blockage, motivating AoA/zenith, Multi-RTT, carrier-phase, barometric, and inertial fusion. Second, TN--NTN systems need joint synchronization, ephemeris-aware weighting, Doppler-aided timing correction, and adaptive node selection. Third, verification requires residual tests, confidence metrics, and integrity monitoring that separate location error from unreliable or manipulated reports; this also requires security hooks against spoofing, jamming, false UE reports, and unauditable AI/ML confidence outputs. Fourth, node selection should reduce overhead without degrading geometry, using CRLB/GDOP criteria or learned reliability rankings from signal quality, residuals, NLoS indicators, and mobility context \cite{Fang2025Lightweight,Sinha2022TOA,Motie2024SelfUAV}. Fifth, benchmarkable TN--NTN datasets should report horizontal and vertical error, latency, signaling load, confidence, integrity risk, and verification success rate \cite{Wymeersch2009Cooperative,Win2011Cooperation}.

\section{Future Outlook and Design Guidelines}
Future 6G positioning should use layered hybrid localization: observables with quality indicators, geometry-aware fusion and residual tests, selective node usage, low-overhead assistance, confidence and integrity reporting, and benchmarks covering horizontal and vertical error, latency, overhead, privacy, and verification.

\section{Conclusion}
This survey synthesized the evolution from LTE and GNSS positioning to NR and NTN-enabled 3D positioning through a standards-aware, observable-oriented lens. LTE established the cellular framework, GNSS remains a key outdoor 3D reference, NR expands the toolbox through timing, angle, carrier phase, sensors, and hybrid operation, and NTN shifts the objective toward wide-area and network-verified location. The main conclusion is that future positioning should be viewed as hybrid TN--NTN fusion with explicit treatment of geometry, vertical observability, reliability, and integrity, rather than as a competition among individual methods. Remaining gaps include trustworthy verification, cross-domain synchronization, reliability-aware node selection, and benchmarkable 3D evaluation aligned with evolving 3GPP procedures.

\section*{Acknowledgment}
This work was supported by the German Federal Ministry of Research, Technology and Space (BMFTR) through Open6GHub+ {16KIS2406}, X-COM {16KISS007K}, and 6G-Coverage {16KIS2425}. The authors are solely responsible for the paper content.

\bibliographystyle{IEEEtran}
\enlargethispage{3\baselineskip}
\bibliography{references}

\end{document}